# Discovery of an isostructural phase transition within orthorhombic phase field of CaTiO$_3$

Saurabh Tripathi[1], Anil Kumar[2], Sharmila Shirodkar[3], Masatomo Yashima[4], U. V. Waghmare[2], Dhananjai Pandey[1]

[1]School of Materials Science and Technology, Indian Institute of Technology (Banaras Hindu University), Varanasi-221005, India

[2]Theoretical Sciences Unit, Jawaharlal Nehru Centre for Advanced Scientific Research, Bangalore 560064, India

[3]Department of Chemistry and Materials Science, Graduate School of Science and Engineering, Tokyo Institute of Technology, 2-12-1-W4-17, O-okayama, Meguro-ku, Tokyo, 152-8551, Japan

Earth's lower mantle extending from 670 to 2,990 km deep is predominantly composed of a perovskite-type (Mg,Fe)SiO$_3$ phase[1,2]. The perovskite phase undergoes a structural phase transition to a post-perovskite phase responsible for D" layer seismic discontinuity[2,3] at about 2690 km depth in the lowermost region of the lower mantle. However, structural basis of other seismic discontinuities occurring in the upper region of the lower mantle (700 km to 1,200 km deep) remains unexplained[4-7], as no apparent change in the crystal symmetry of the orthorhombic perovskite phase has been reported[5]. We present here unambiguous evidence for a non-apparent isostructural phase transition[8] in the stable orthorhombic perovskite phase of CaTiO$_3$ which may have relevance to phase transitions in the perovskite phase of (Mg,Fe)SiO$_3$ also, as both the compounds have similar structure, tolerance factor and thermochemical properties[9-11]. Our results are based on the analysis of neutron powder diffraction patterns using Rietveld and mode crystallography techniques and are supported by density functional and Landau theory calculations. The present results on CaTiO$_3$ would encourage search for isostructural phase transition in the perovskite phase of (Mg,Fe)SiO$_3$ that may provide clue to the unexplained geophysical phenomena in the upper part of the earth's lower mantle.

The family of oxide perovskites with the general formula $ABO_3$ is named after the mineral name of $CaTiO_3$[9-10] where $A$ and $B$ are larger and smaller cations, respectively. $CaTiO_3$, whose structure and the tolerance factor are similar to the perovskite phase of the earth's lower mantle material $(Mg,Fe)SiO_3$[10-11], undergoes structural phase transitions at easily achievable temperatures[12], as compared to $(Mg,Fe)SiO_3$ that requires extremely high temperatures and pressures[2]. $CaTiO_3$ has therefore been used as a model system to mimic the phase transition behaviour of $(Mg,Fe)SiO_3$. The stable high temperature cubic phase (space group Pm3m) of $CaTiO_3$ transforms to a tetragonal phase (space group *I4/mcm*) at $T \sim 1636$ K which can be visualized to occur through condensation of one of the components of the $R_{4+}$ mode at $\boldsymbol{q}$= ½ ½ ½ of the cubic Brillouin zone culminating in anti-phase rotated oxygen octahedra in the $a^0 a^0 c^-$ tilt system[12-13]. This tetragonal phase on further cooling below $T = 1512$ K transforms to an orthorhombic phase in the *Pnma* space group with $a^- b^+ a^-$ tilt system[14-15] due to the condensation of $M_{3+}$ mode[12-13] at q= ½ ½ 0 involving in-phase rotation and another component of the $R_{4+}$ mode causing additional anti-phase rotation of the octahedra. This orthorhombic phase is widely accepted as the stable phase of $CaTiO_3$[12-13] under ambient conditions and of $MgSiO_3$ under the lower mantle conditions[4]. From an analysis of the temperature dependent neutron diffraction patterns of $CaTiO_3$ within the stability field of the orthorhombic Pnma phase, we present evidence for discontinuous changes in the unit cell volume, pseudomonoclinic cell distortion angle and the in-phase as well as anti-phase tilt angles in between 1000-1069K indicative of a first order isostructural phase transition that does not involve any change of space group and/or Wyckoff site symmetries. We show that this transition is phonon driven as the amplitudes of the $R_{4+}$ and $M_{3+}$ phonon modes, obtained from amplimode analysis of neutron powder diffraction data and first principles calculations, exhibit anomalous non-analytic behavior at the transition temperature. We also present results of Landau theory to show how a synergy between the two order parameters emerging at higher temperature phase transitions involving in-phase and anti-phase tilts of oxygen octahedra can lead to such an isostructural phase transition.

In $CaTiO_3$[12], as also in $MgSiO_3$[16], the hhh reflections are doublet and hence the equivalent elementary perovskite cell can be taken to be pseudomonoclinic (for more details, see the structure refinement section of the supplementary file). The pseudomonoclinic cell parameters of $CaTiO_3$ at room temperature show the relationship[17] : $a_{pm}=c_{pm}<b_{pm}$, $\alpha=\gamma=90^0\neq\beta$ . The temperature variation of the elementary perovskite cell parameters $a_{pm}=c_{pm}$ , $b_{pm}$ and the monoclinic angle $\beta_{pm}$ of $CaTiO_3$, as obtained from Rietveld refinements, shown in Fig 1 reveals that the relationship $a_{pm}=c_{pm}<b_{pm}$ changes to $a_{pm}=c_{pm}>b_{pm}$ somewhere between 1000 to 1069K. Further, the monoclinic angle $\gamma_{pm}$ changes discontinuously in between 1000-1069K (see inset 'a' in Fig. 1). The unit cell volume also shows a discontinuous change in the same temperature range (see inset 'b' in Fig. 1). These discontinuous changes in the unit cell parameters occur within the stability field of the orthorhombic phase in the Pnma space group, below the cubic to tetragonal and tetragonal to orthorhombic phase transitions at 1636 and 1512K, respectively, and therefore represent an isostructural phase transition (ISPT), which is quite rare in nature[9,18,19].

In the $a^-b^+a^-$ tilt system of the Pnma phase of CaTiO$_3$, the tilt angles for the anti-phase rotations (a$^-$) of the oxygen octahedra about the pseudocubic [100]$_p$ and [001]$_p$ axes are equal and are equivalent to a single tilt through an angle φ about one of the two diad axes parallel to the [101]$_p$ and [10$\bar{1}$]$_p$ directions, where the subscript p refers to the ideal high symmetry elementary perovskite cell. The in-phase octahedral rotation (b$^+$) through an angle ζ is about the [010]$_p$ axis. Both the tilt angles, φ and ζ, can be determined from the coordinates of the apical oxygen O$_I$, which changes from 0.25, 0, 0.75 in the cubic phase to 0.25+u, v, 0.75-w in the orthorhombic structure, using the relationships given in Ref [20-21]: It is evident from Fig 2 (a, b), which depicts the temperature variation of the anti-phase (φ) and in-phase (ζ) tilt angles, that both the angles undergo a discontinuous change in between 1000-1069K similar to the unit cell parameters shown in Fig 1. Since the tilt angles are the macroscopic order parameters for the AFD phase transition, their discontinuous change in between 1000-1069K in Fig 2 (a, b) confirms the existence of a first order AFD transition of isostructural type. To understand the likely role of phonons in the AFD isostructural transition, we also determined the amplitudes of the R$_{4+}$ and M$_{3+}$ modes of the cubic perovskite structure[22-23] from the refined positional coordinates using the Amplimodes suite[24-25]. We find that the amplitudes of R$_{4+}$ and M$_{3+}$ modes change discontinuously at the isostructural phase transition (see Fig 2 (a, b). This suggests the role of phonons in this transition, as confirmed further using first principles calculations.

For a system with more than one unstable mode, there could be competition as well as coupling between the modes to drive the system to the most stable structure. The amplitude of the unstable mode, which is a measure of deformation of the high symmetric high temperature structure that yields the ground state structure, provides important information about the structural phase transitions. For a given structure, the amplitude of unstable mode shows a continuous behaviour as a function of temperature or external field, but shows a non-analytic behaviour close to a phase transition. The unstable modes at M and R points in the Brillouin zone of cubic $CaTiO_3$ "freeze-in" or distort the structure leading to the orthorhombic phase at low temperatures. Our calculations of the cubic structure show that the M and R modes are unstable in $CaTiO_3$, in agreement with the findings of previous workers[26-27], and their frequency decreases monotonically with temperature as shown in Fig.3 (a). To understand the behaviour of M and R mode amplitudes and associated phase transitions with temperature, as observed in our experiments, we allow the system to relax to minimum energy by freezing the unstable zone boundary phonon modes using a $\sqrt{2}$ x $\sqrt{2}$ x 2 supercell. The energy profile as a function of atomic displacements along the soft eigen modes provides information about the mode amplitude (of atomic displacements) in the minimum energy structure. We determined energy profile as a function of atomic displacements for the experimental structural parameters at different T (see Fig. 2 (a, b) of Extended data figure file). We find that energy gain by freezing of the $M_{3+}$ and $R_{4+}$ modes decreases with the increase of temperature (a similar trend can be seen in the increase of phonon frequencies for these modes with temperature in Fig 3(a)). The amplitudes of $M_{3+}$ and $R_{4+}$ modes decrease continuously with increasing temperature but show a non-analytical behaviour with a discontinuous change in between 1000 to 1069K (see Fig. 3(a)), consistent with our experimental observations shown in Fig. 2(a, b). Above the isostructural phase transition temperature, the temperature dependence of the two amplitudes are continuous within the stability limit of the Pnma phase.

In complement to first-principles demonstration of the discontinuity in $R_{4+}$ and $M_{3+}$ phonon modes across the isostructural transition occurring somewhere in between 1000 to 1069K (as a result of observed discontinuity in volume), we now use phenomenological Landau theory and show that this isostructural phase transition originates from the synergistic coupling between order parameters emerging at higher temperature transitions. From the Landau free energy of higher temperature phase transitions (see Supplementary file for complete derivation), we obtain free energy as a function of x and y, the two order parameters related to the $R_{4+}$ mode (along (0,a,a)) and $M_{3+}$ mode (along (a,0,0)), respectively[28].

$$F = e_1 x + \frac{a_1}{2}(\frac{T}{T_C} + p)x^2 + \frac{b_1}{3}x^3 + \frac{c_1}{4}x^4$$
$$+ e_2 y + \frac{a_2}{2}(\frac{T}{T_H} + p)y^2 + \frac{b_2}{3}y^3 + \frac{c_2}{4}y^4 \qquad (1)$$
$$+ d_1 x^2 y + d_2 xy^2 + d_3 xy + d_4 x^2 y^2,$$

where {$e_1$, $a_1$, $b_1$, $c_1$} are the Landau coefficients for the order parameter (x) associated with the $R_{4+}$ mode (along (0,a,a)), and {$e_2$, $a_2$, $b_2$, $c_2$} are the Landau coefficients of the order parameter (y) associated with $M_{3+}$ mode [along (a,0,0)], and $d_i$'s (i=1 to 4) are the coupling constants. Renormalization of the coefficients of the quadratic and linear terms (refer to Supplementary file) gives the effective transition temperatures of $T_C$ and $T_H$. We fit them to 1000 K to reproduce the experimentally observed isostructural phase transition (refer to Supplementary file). The values of parameters were fixed at $e_1$= 0.3,

$a_1=1$, $b_1=2$, $c_1=1$, $T_C=1000$ K and p= 0, and $e_2=0.1$, $a_2=1$, $b_2=2$, $c_2=1$, $T_H=1000$ K and q= 0, such that the value of free energy functions of the decoupled $x$ and $y$ order parameters is 0 at $T_C$ and $T_H$, respectively. The coupling coefficients $d_1$, $d_2$, $d_3$, $d_4$ are 0.3, 0.1, 0.4 and 0.1 respectively. Equilibrium values of $x$ and $y$ are obtained through minimization of free energy at each temperature[29] and are depicted in Fig. 3 (b), which shows a clear discontinuity in both the $x$ and $y$ order parameters near 1000K, in agreement with experimental observations. Thus, the isostructural phase transition is a result of nonlinear coupling between two distinct three-component order parameters that have already broken the symmetry, and the richness in the symmetry equivalent structures possible.

We believe that the discovery of the isostructural phase transition in $CaTiO_3$ will encourage search for a similar transition within the stability field of the orthorhombic perovskite phase of $MgSiO_3/(Mg,Fe)SiO_3$ that may hold clue to the hitherto unexplained geophysical phenomena[5] in the upper part of the earth's lower mantle consisting of perovskite phase of $(Mg,Fe)SiO_3$ predominantly.

**Methods:** The details of sample preparation, neutron data collection, Rietveld refinements and Amplimode analysis, first principles calculations and Landau theory are given in the supplementary files.

**Supplementary Information** is linked to the online version of the paper at www.nature.com/nature.

**Author Contributions:** D. P. conceived the problem and applied the concept of isostructural phase transition, supervised the analysis of the neutron diffraction patterns as also the theoretical calculations and prepared the final version of the manuscript. ST analysed the neutron diffraction patterns and prepared the first draft of the paper. MY provided the neutron diffraction data collected on a sample prepared in his laboratory. The details of sample preparation are given in Ref. No. 12. The data used in Ref. No. 12 was reanalyzed by ST leading to the discovery of the isostructural phase transition in the orthorhombic phase of $CaTiO_3$. AK and UW carried out first principle calculations while SS and UW carried out the Landau theory calculations. AK, SS and UW prepared the first draft of the theoretical portion of the manuscript.

**Figure Legends**

Fig. 1 Variation of equivalent elementary perovskite cell parameters $a_{pm} = c_{pm}$ , $b_{pm}$ with temperature. Inset shows variation of (a) monoclinic angle $\beta_{pm}$ and (b) volume of $CaTiO_3$ with temperature. Error bars are displayed in the figures.

Fig. 2 Variation of (a) inphase tilt angle ($\zeta$) and amplitude of $M_{3+}(A)$ primary mode, (b) antiphase tilt ($\varphi$) and amplitude of $R_{4+}(A)$ primary mode of Pnma phase of $CaTiO_3$ with temperature. Error bars for tilt angles are displayed in the figures.

Fig. 3 (a) Variation of frequency ($\omega$) and amplitude (A) of soft R and M primary modes for cubic structure of $CaTiO_3$ with temperature, (b) 'x' and 'y' order parameters with temperature[29]. Note that since the values of 'x' and 'y' are negative, here we have plotted the absolute values.

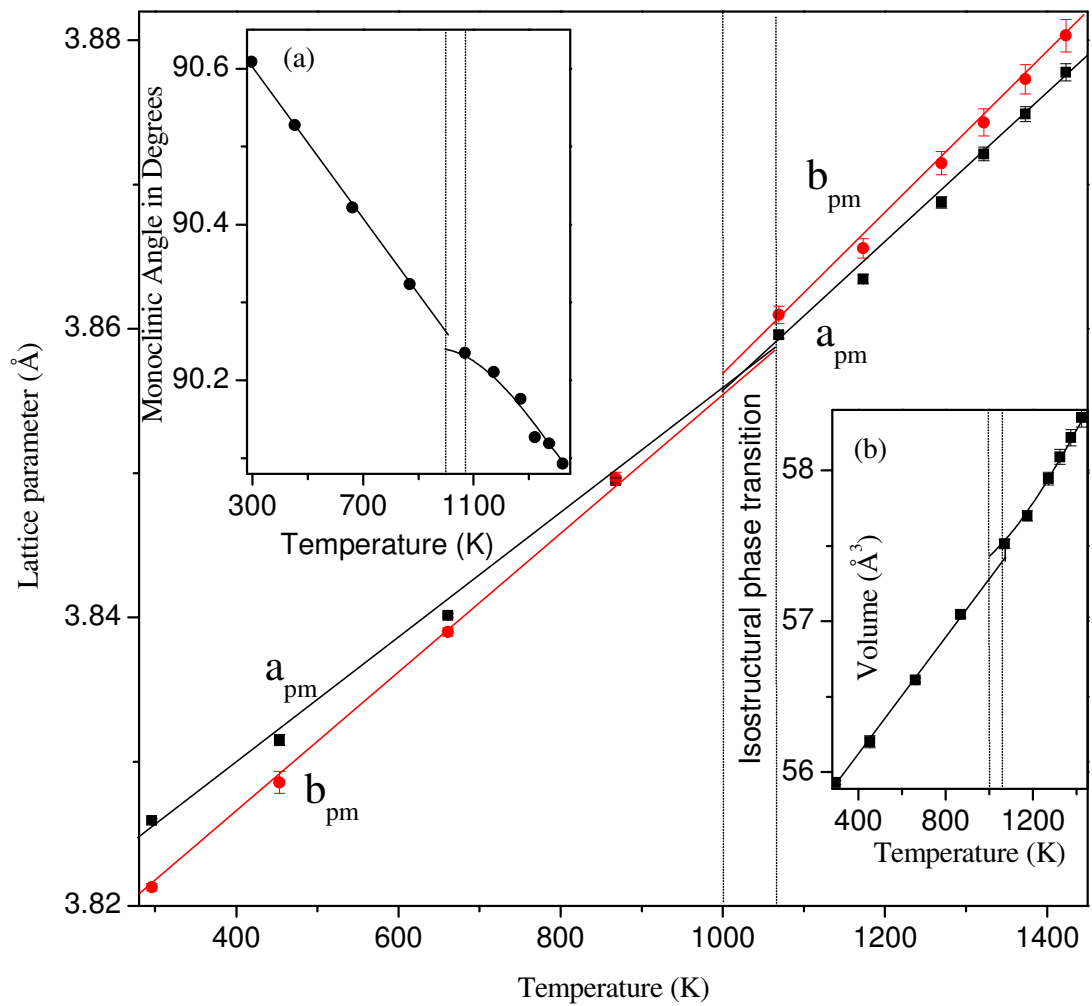

Fig. 1

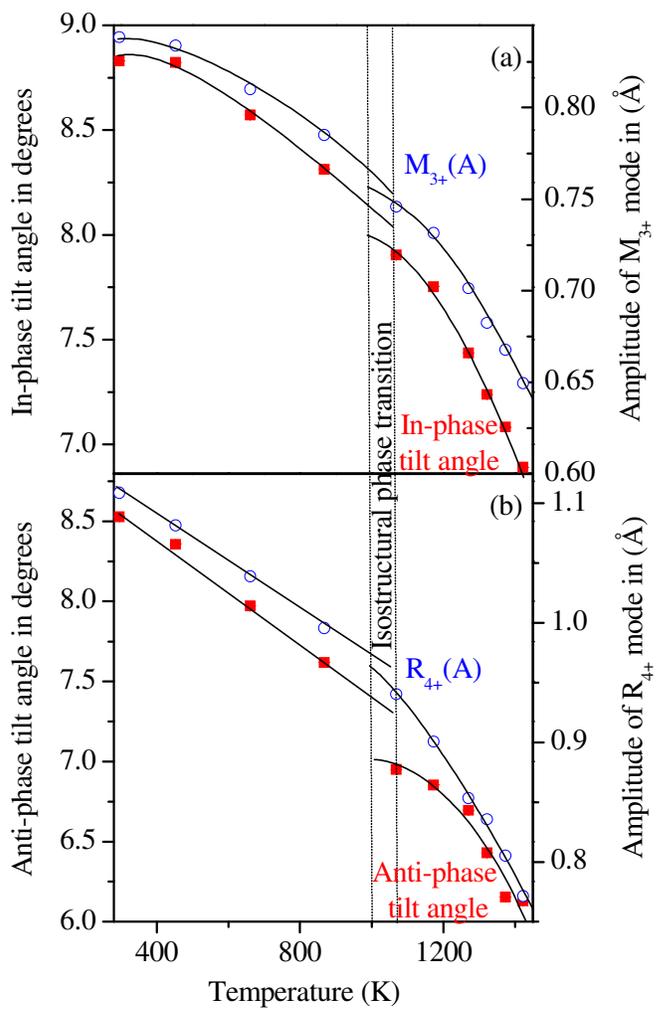

Fig. 2

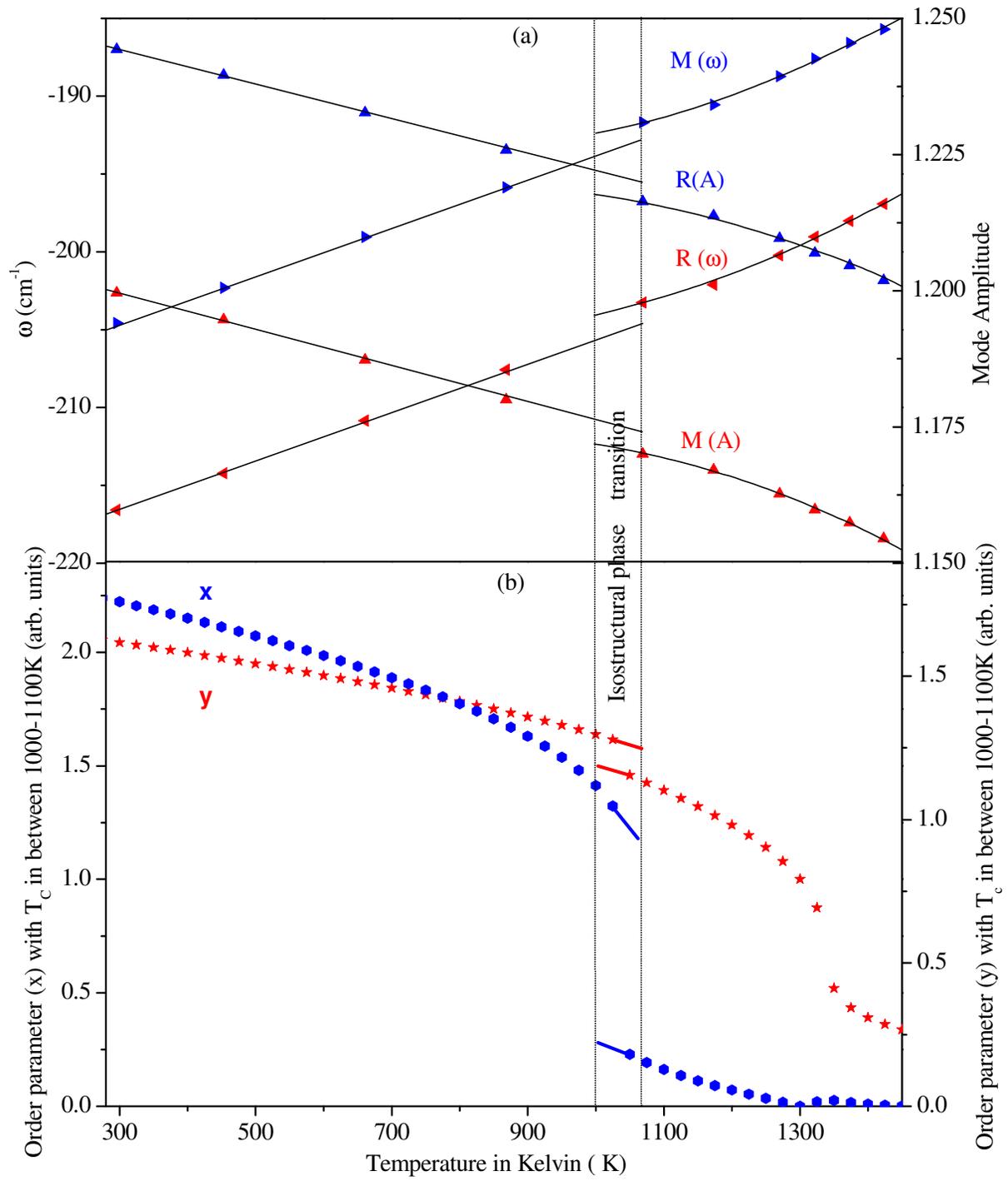

Fig. 3

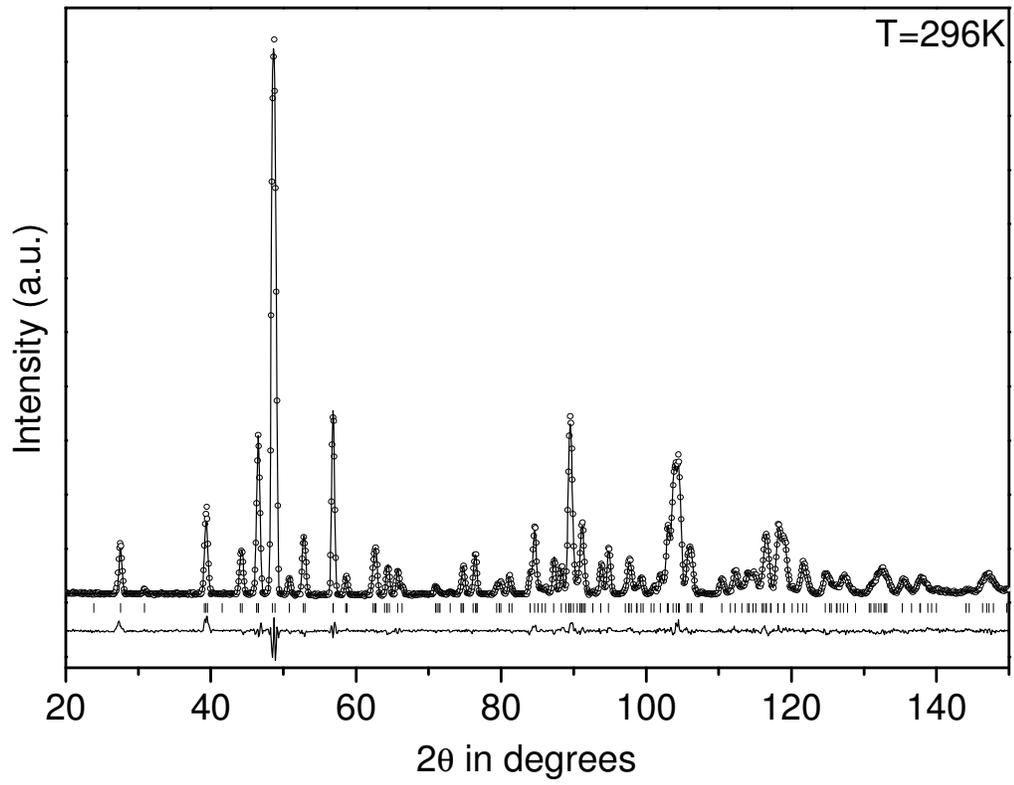

T=296K

Intensity (a.u.)

20    40    60    80    100    120    140

2θ in degrees

Fig. 1(a)

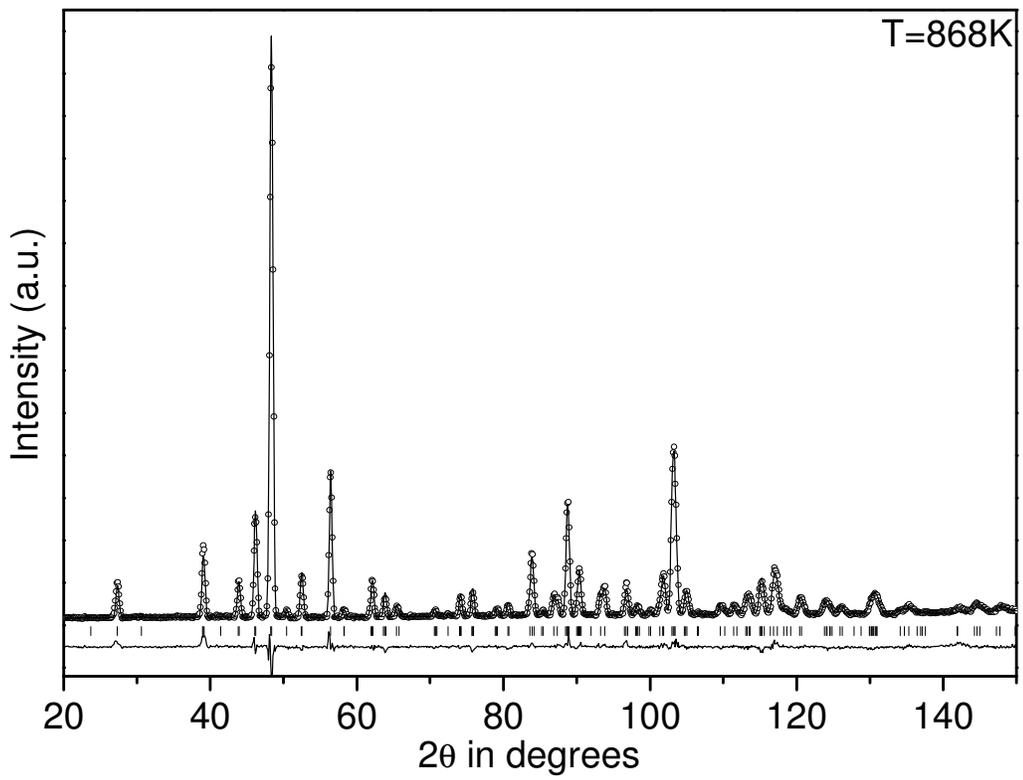

T=868K

Intensity (a.u.)

20    40    60    80    100    120    140

2θ in degrees

Fig. 1(b)

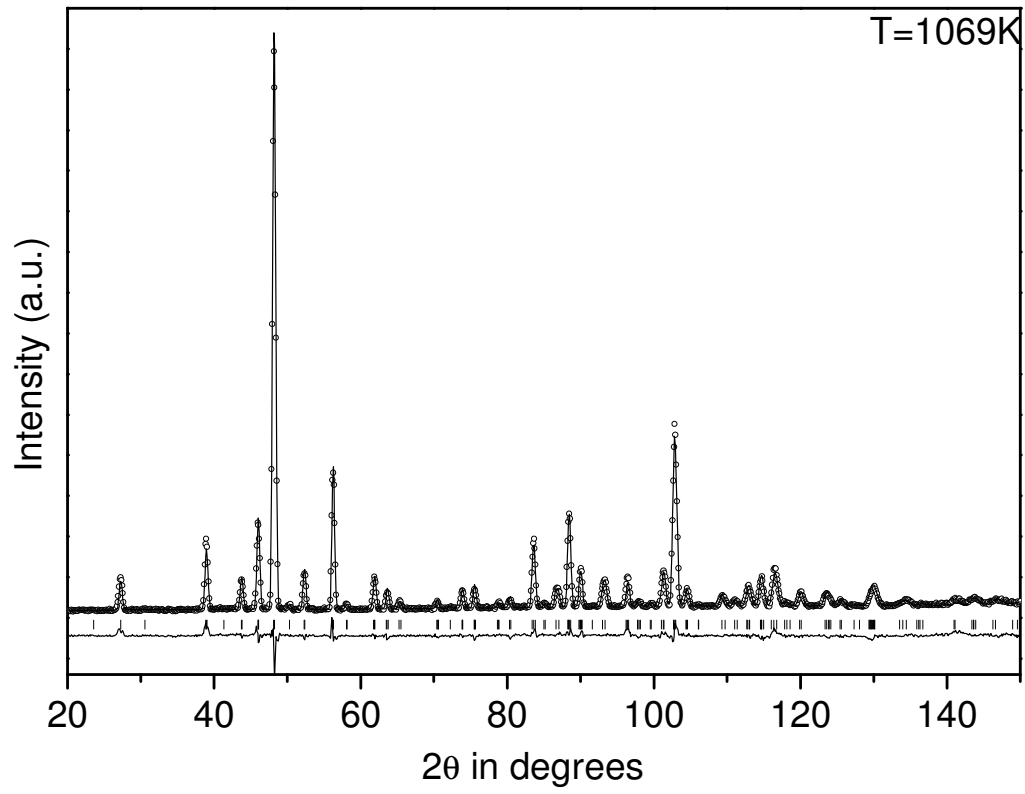

T=1069K

Intensity (a.u.)

2θ in degrees

Fig. 1(c)

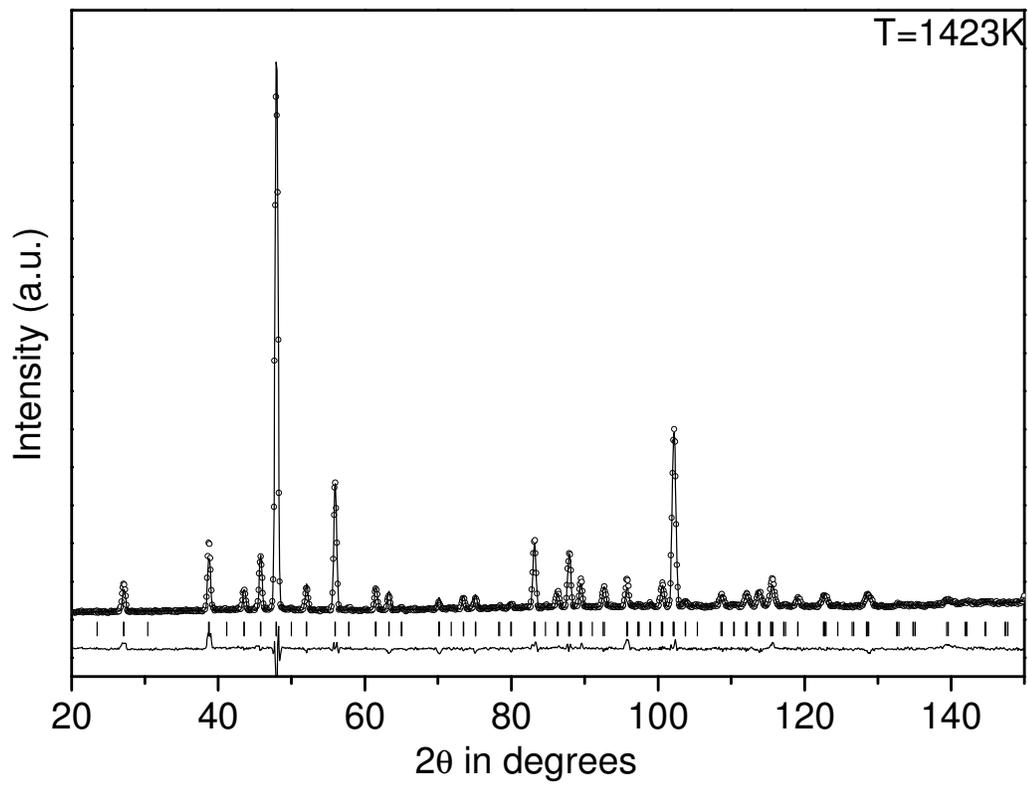

T=1423K

Intensity (a.u.)

2θ in degrees

Fig. 1(d)

Extended data Fig. 1 (a, b, c, and d) depicts the observed, calculated and difference profiles obtained after final cycle of refinement for T= 296K, 868K, 1069K and 1423K using Pnma space group.

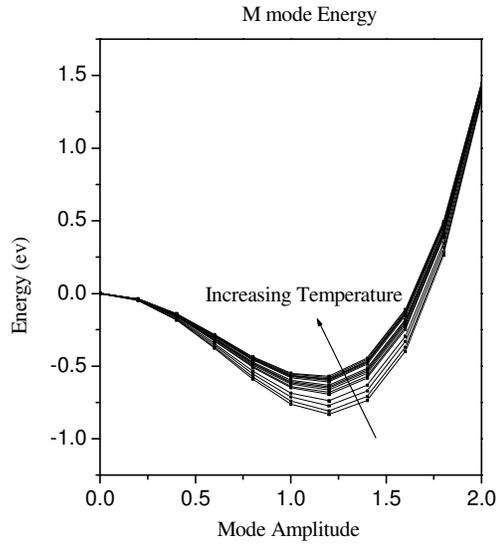

Fig. 2 (a)

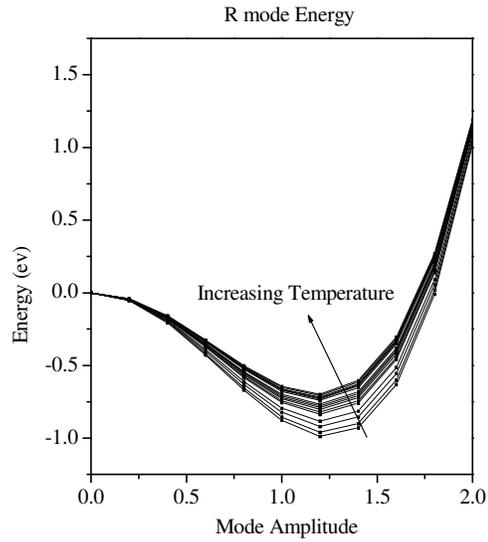

Fig. 2 (b)

Extended data Fig. 2 (a, b) Energy profile as a function of atomic displacements for the experimental structural parameters at different Temperature.

## Supplementary Information

### Structure refinement

The Rietveld refinements using the neutron powder diffraction patterns of $CaTiO_3$ were performed by FULLPROF suite[1]. The complete Fullprof suite and documentation is available at the web[2]. The peak shape was assumed to be pseudo-Voigt function with asymmetry. The background of each profile was approximated by a 6 coefficient polynomial. Except for the occupancy parameters of the atoms, which were kept at their nominal composition, all other parameters, i.e., scale factor, zero correction, background, half width parameters along with mixing parameters, lattice parameters, positional coordinates, and thermal parameters, were refined.

Amplitudes of phonon modes were calculated using AMPLIMODES package from Bilbao Crystallographic Server[3,4].

The asymmetric unit of the orthorhombic phase of $CaTiO_3$ in the Pnma space group consists of 4 atoms, viz Ca, Ti, O1 (apical oxygen) and O2 (planar oxygen) as

| | | | | | |
|---|---|---|---|---|---|
| Ca | 1 | 4c | 0.0+/-u | 0.25 | 0+/-w |
| Ti | 1 | 4b | 0.0 | 0.0 | 0.5 |
| O | 1 | 8d | 0.25+/-u | 0.0+/-v | 0.75+/-w |
| O | 2 | 4c | 0.50+/-u | 0.75 | 0+/-w |

There are 7 refineable atomic xyz coordinates. The coordinates of all the atoms in the asymmetric unit at various temperatures from 296K to 1423K were obtained by Rietveld refinements. Fig. 1 (a, b, c, and d) of the Extended data figure file, depicts the observed, calculated and difference profiles obtained after final cycle of refinement for T= 296K, 868K, 1069K and 1423K using Pnma space group. Evidently, all the fits are very good as can be seen from the near flatness of the difference profile.

In order to explain the isostructural phase transition we have expressed orthorhombic structure (supercell) of $CaTiO_3$ with Pnma space group having $a^-b^+a^-$ tilt system into pseudomonoclinic elementary perovskite cell (subcell). The relationships between the orthorhombic cell parameters $A_0$, $B_0$ and $C_0$ and the elementary pseudocubic perovskite cell parameters $a_{pm}$, $b_{pm}$, and $c_{pm}$ are : $(a_{pm})^2 = (c_{pm})^2 = (A_0/2)^2 + (C_0/2)^2$, $A_0^2 = a_{pm}^2 + c_{pm}^2 - 2a_{pm}c_{pm}\cos\beta$, and $C_0^2 = a_{pm}^2 + c_{pm}^2 + 2a_{pm}c_{pm}\cos\beta$, and $B_0 = 2b_{pm}$, where $\beta_{pm}$ is the angle between the $a_{pm}$ and $c_{pm}$ axes. The other two angles are $\alpha_{pm} = \gamma_{pm} = 90^0 \neq \beta_{pm}$ making the elementary cell pseudomonoclinic. For $A_0 = C_0$, the elementary perovskite cell axes are orthogonal to each other. The hhh type reflections are doublet and singlet for the monoclinic and orthogonal cells, respectively. In $CaTiO_3$, as also in $MgSiO_3$, the hhh reflections are doublet and hence the equivalent elementary perovskite cell can be taken to be pseudomonoclinic.

**Mode Crystallography:**

Mode crystallographic analysis was done using AMPLIMODE suite at Bilbao Crystallographic Server[3,4], using Rietveld refined coordinates. AMPLIMODES is a computer program which calculates the amplitude of symmetry adapted modes for the low symmetry distorted structure with respect to a high symmetry structure. These modes are given in relative units using atomic displacements of the atoms in the asymmetric unit of the distorted phase.

In order to work with the AMPLIMODE analysis, we need to know the parent [or high symmetry (HS)] structure, the [low symmetry structure (LS)] and the transformation matrix for the two phases, and the symmetry adapted primary and secondary modes that transform the HS phase into the LS phase. In the case of $CaTiO_3$, the high symmetry structure is cubic with space group $Pm\bar{3}m$, where A site cation, B site cation and O go to 1b, 1a and 3d Wyckoff sites, with fractional coordinates 0.5 0.5 0.5, 0.0 0.0 0.0 and 0.5 0.0 0.0, respectively.

The transformation matrices relating the unit cell of the low symmetry room temperature phase in the Pnma space group with the unit cell of the high symmetry cubic phase of $CaTiO_3$ is given below:

$$\begin{bmatrix} 1 & 0 & 1 \\ 0 & 2 & 0 \\ -1 & 0 & 1 \end{bmatrix} \begin{bmatrix} 1/2 \\ 0 \\ 1/2 \end{bmatrix}$$

The Wyckoff sites of the cubic phase get split via condensation of various symmetry adapted modes to give the Pnma space group for the room temperature structures of $CaTiO_3$. It has been shown that seven modes are required to arrive at the Pnma structure from the $Pm\bar{3}m$ structure[5,6]. The irreps corresponding to these seven modes are: octahedral tilting modes R4+ (anti-phase tilt) and M3+ (in-phase tilt), octahedral distortion modes X5+, R5+ and M2+, and A-site cation displacement modes R5+ and X5+[7]. Using the package 'Symmodes' at the Bilbao Crystallographic server[8], we find that the displacive phase transition from cubic space group $Pm\bar{3}m$ to the orthorhombic space group Pnma requires the splitting of the cubic Wyckoff sites through these vibrational modes in the following manner:

(i)     Ca at 1b site of $Pm\bar{3}m$ goes to 4c site via R5+(1) and X5+(1) symmodes with coordinates as (0.00+u, 0.25, 0.00-w),

(ii)    O at 3d site goes to O1 at 8d (0.25+u, 0.00+v, 0.75-w) and O2 at 4c (0.50+u, 0.75, 0.00-w) sites via R4+(1), R5+(1), X5+(1), M2+(1), and M3+(1) symmodes and

(iii)   Ti at 1a site remains undisplaced from its position and occupies 4b site with coordinates (0.00, 0.00, 0.50),

The above representations of irreps are obtained from Bilbao crystallographic server. Thus this process gives us amplitudes of phonon modes using Rietveld refined coordinates of low symmetry structure.

**First principle calculations**

Density functional theory calculations were performed using ABINIT code[9] and local density approximation (LDA) for the exchange-correlation functional. We used norm conserving pseudopotentials with an energy cutoff 40 Ha and $6 \times 6 \times 6$ Monkhorst-Pack mesh for the sampling of the cubic Brillouin zone. We approximate the experimental lattice constants of cubic structure for the different values of temperature (T) from the volume of orthorhombic unit cells and performed calculations using these approximated cubic lattice constants. The phonon frequencies are the eigen values of the dynamical matrix D:

$$D_{i\alpha j\beta} = (1/\sqrt{m_i\, m_j})x\ (\partial^2 E/\partial x_{i\alpha}\partial x_{j\beta}) \qquad\qquad (1)$$

where $m_i$ is the mass of atom i, $x_{i\alpha}$ is the displacement of atom i along cartesian direction $\alpha$ and E is total energy of the system. The phonon dispersion relations and soft mode eigen vectors are obtained from the linear response calculations in a cubic unit cell. The phonon frequencies are proportional to the curvature of energy surface as a function of small displacement of atoms from their equilibrium positions (see Eq. 1). The unstable modes, which correspond to the negative curvature of the energy surface, are a signature of the structural phase transitions. The presence of unstable phonon modes shows that system can lower its energy from such displacement patterns, which also lowers the symmetry of structure leading to the structural phase transitions. The more is the curvature of energy surface for a unstable mode, larger is the probability of system following that displacement pattern.

**Phenomenological theory**

$R_{4+}$ and $M_{3+}$ are the two modes that lead to the transition of cubic Pm-3m phase to the Pnma phase. $R_{4+}$ and $M_{3+}$ modes correspond to tilting of octahedra, and are triply degenerate. Let $\{x_1, x_2, x_3\}$ and $\{y_1, y_2, y_3\}$ be the order parameters associated with the $R_{4+}$ and $M_{3+}$ modes, respectively. The free energy functional, F(Pm-3m) with Pm-3m as the parent structure and all the six order paramters is given by,

$$F(Pm-3m) = g_{12}(T)(x_1{}^2 + x_2{}^2 + x_3{}^2) + g_{14}(x_1{}^2 + x_2{}^2 + x_3{}^2)^2 + g_{24}(x_1{}^4 + x_2{}^4 + x_3{}^4)$$
$$+ f_{12}(T)(y_1{}^2 + y_2{}^2 + y_3{}^2) + f_{14}(y_1{}^2 + y_2{}^2 + y_3{}^2)^2 + f_{24}(y_1{}^4 + y_2{}^4 + y_3{}^4) \qquad (S1)$$
$$+ h_{14}(x_1{}^2 + x_2{}^2 + x_3{}^2)(y_1{}^2 + y_2{}^2 + y_3{}^2) + h_{24}(x_1{}^2 y_1{}^2 + x_2{}^2 y_2{}^2 + x_3{}^2 y_3{}^2).$$

Freezing in of the $R_{4+}$ mode along (a,0,0) gives the I4/mcm phase. Hence, substituting $x_1 = x_c$ (constant) and neglecting the constant terms in equation S1 we get,

$$F(I4/mcm) = (g_{12}(T) + 2g_{14}x_c{}^2)(x_2{}^2 + x_3{}^2) + g_{14}(x_2{}^2 + x_3{}^2)^2 + g_{24}(x_2{}^4 + x_3{}^4)$$
$$+ (f_{12}(T) + 2h_{14}x_c{}^2)(y_1{}^2 + y_2{}^2 + y_3{}^2) + f_{14}(y_1{}^2 + y_2{}^2 + y_3{}^2)^2 + f_{24}(y_1{}^4 + y_2{}^4 + y_3{}^4) \qquad (S2)$$
$$+ h_{14}(x_2{}^2 + x_3{}^2)(y_1{}^2 + y_2{}^2 + y_3{}^2) + h_{24}(x_c{}^2 y_1{}^2 + x_2{}^2 y_2{}^2 + x_3{}^2 y_3{}^2).$$

Below 1512 K, the $R_{4+}$ and $M_{3+}$ modes freeze into the tetragonal I4/mcm structure along (0,a,a) and (a,0,0) respectively, and give rise to the orthorhombic Pnma phase. Thus, the temperature dependence enters F through $g_{12}(T)= (T/1512)-1$ and $f_{12}(T)= (T/1512)-1$. Since the isostructural phase transition involves freezing in of both $R_{4+}$ and $M_{3+}$ modes along (0,a,a) and (a,0,0) directions respectively, the order parameters x and y are variable for T <1512 K. Hence substituting $x_2=x_3= x + x_o$, and $y1= y + y_o$ and $y_2=y_3=0$ ($x_o$ and $y_o$ are constants), equation S2 reduces to,

$$
\begin{aligned}
F(Pnma) = & [4x_o(g_{12}(T) + 2g_{14}x_c^2) + 4x_o^3(4g_{14} + 2g_{24}) + 4h_{14}x_o y_o]x \\
& + [2g_{12}(T) + 2g_{14}x_c^2 + 6x_o^2(2g_{14} + 2g_{24} + h14 y_o^2)]x^2 \\
& + [4x_o(g_{14} + 2g_{24})]x^3 + (4g_{14} + 2g_{24})x^4 \\
& + [(f_{14} + f_{24})4y_o^3 + (f_{12}(T) + h_{14}x_c^2)^2 y_o + 2h_{24}x_c^2 y_o + 4h_{14}x_o^3]y \\
& + [f_{12}(T) + h_{14}x_c^2 + 6(f_{14} + f_{24})y_o^2 + 2h_{14}x_o^2]y^2 + [4y_o(f_{14} + f_{24})]y^3 \\
& + (f_{14} + f_{24})y^4 \\
& + (8h_{14}x_o y_o)xy + (4h_{14}y_o)x^2 y + (4h_{14}x_o)xy^2 + 2h_{12}x^2 y^2
\end{aligned}
\tag{S3}
$$

Redefining the Landau coefficients we get,

$$
\begin{aligned}
F = & e_1 x + \frac{a_1}{2}\left(\frac{T}{T_C} + p\right)x^2 + \frac{b_1}{3}x^3 + \frac{c_1}{4}x^4 \\
& + e_2 y + \frac{a_2}{2}\left(\frac{T}{T_H} + p\right)y^2 + \frac{b_2}{3}y^3 + \frac{c_2}{4}y^4 \\
& + d_1 x^2 y + d_2 xy^2 + d_3 xy + d_4 x^2 y^2,
\end{aligned}
\tag{S4}
$$

Since the I4/mcm to Pnma phase transition is relevant to equation S4, the critical temperature should be $T_C$ = 1512 K. However all the Landau coefficients are renormalized, and hence the temperature at which x and y order parameters will freeze into the structure is no longer 1512 K. The critical temperature is now replaced by effective temperature, which is $T_C= T_H= 1000$ K. Note that, we fit the $T_C= T_H$ values to 1000 K to mimic the isostructural phase transition in between 1000-1100K.